\documentclass[twocolumn,showpacs]{revtex4}
\usepackage{graphicx}

\begin{document}

\title{Foundations of Quantum Discord}
\author{Vlatko Vedral}
\affiliation{Clarendon Laboratory, University of Oxford, Parks Road, Oxford OX1 3PU, United Kingdom\\Centre for Quantum Technologies, National University of Singapore, 3 Science Drive 2, Singapore 117543\\
Department of Physics, National University of Singapore, 2 Science Drive 3, Singapore 117542\\
Center for Quantum Information, Institute for Interdisciplinary
Information Sciences, Tsinghua University, Beijing, 100084, China}

\begin{abstract}
This paper summarizes the basics of the notion of quantum discord and how it relates to other types of correlations in quantum physics. We take the fundamental information theoretic approach and illustrate our exposition with a number of simple examples. 
\end{abstract}

\maketitle

\section{Introduction}

In order to understand quantum discord we need to first understand the difference between classical and quantum correlations as quantified by mutual information. 
Mutual information is originally a classical measure of correlations. It is defined as
\begin{equation}
I (A,B) = H(A) + H(B) - H(A,B)
\end{equation}
where $S=-\sum_n p_n \log p_n$ is the Shannon entropy \cite{Shannon} and $A$ and $B$ are two random variables whose probability distribution is given. There is a different way of writing this quantity using the conditional entropy, $H(A/B) = H(A,B) - H(B)$, namely,
\begin{equation}
I(A,B) = H(A) - H(A/B)
\end{equation}
Therefore, there are two equivalent ways of thinking about classical correlations (there are many more, but they are not necesarily relevant to our topic). One is that they are measured by the difference in the sum of local entropies and the total entropy and the other one is that they tell us by how much we can reduce the entropy of one random variable by measuring the other. 

When generalizing the concept of mutual information to quantum physics, it is straightforward to do it using the first expression. All we need do is use the von Neumann instead of the Shannon entropy. The quantum mutual information is defined as
\begin{equation}
I_Q = S(A) + S(B) - S(A,B)
\end{equation}
where $S(A)= - tr \rho_A \log \rho_A$ and the subscript $Q$ just indicates that this is a quantum measure. The second expression for the mutual information, involving the conditional entropy, is, however, harder to upgrade to quantum physics. The reason is that if we do the same substitution of the Shannon with the von Neumann entropy, the resulting entity
\begin{equation}
S(A,B) -  S(B)
\end{equation}
can actually be negative for bipartite quantum systems. Take any pure entangled state of two subsystems (such as $a|00\rangle + b|11\rangle$) and its total entropy vanishes, while the reduced entropy is non-zero. It is therefore hard to call this quantum conditional entropy since the quantity can be negative (and given that entropy quantifies disorder it is hard to see how disorder can be smaller than zero). 

A way out of this is to define the quantum conditional entropy S(A/B) as the average entropy $\sum_n p_n S(\rho_{A,n})$ of states of $A$ after a measurement is made on $B$. There are infinitely many measurements we can perform on $B$, so we will choose the one that makes $S(A/B)$ minimum (we want to learn as much about $A$ by measuring $B$). It is clear that the upper bound on this quantum conditional entropy is $S(A)$, while the lower bound is zero. 

So $S(A/B)$ is always positive when defined as the average entropy reduction, but now we have another problem. The quantum mutual information defined above is not equal to $S(A) - S(A/B)$! Unlike in classical information, the two ways of expressing quantum mutual information are actually different. 

This is because the quantum mutual information can actually reach the value of $2S(A)$, while the quantity to $S(A) - S(A/B)$ can at most be $S(A)$. 
What does the difference between two quantum quantities ($I_Q$ and $S(A) - S(A/B)$) signify, if anything? 

This question was first asked by Lindblad \cite{Lindblad} (he phrased is slightly differently but that was the spirit). His answer was that the difference is actually due to quantum entanglement (more precisely, he says: ``This extra correlation is
of course the cause of the Einstein-Podolsky-Roscn "paradox" and is
thus a typical quantum effect."). And for pure states he was perfectly correct (as we will see below), but at that time there was virtually no work on mixed state entanglement (which only proerly took off in the mid-nineties) and so it was difficult for him to anticipate many subtleties involved.  

\section{Discord: what it is and how to quantify it}

I came upon this difference of mutual informations after my initial work on entanglement because I was asking if the quantum mutual information, which quantifies all correlations, can actually be written as a sum of entanglement and classical correlations. I defined classical correlations as $C(A,B) = S(A/B)$ and I quantified entanglement using the relative entropy of entanglement $E(A,B)$. 

My then student, Leah Henderson, and I discovered that the sum $C + E$ is mostly smaller than $I_Q$ for mixed states \cite{Henderson}. In other words, there is more to quantum correlations than just entanglement when it comes to mixed states. For pure states, entanglement and classical correlations are equal to one another and the sum is then exactly equal to the quantum mutual information which explains why quantum mutual information is twice as big as the classical mutual information (as anticipated by Lindblad). 

A few months later, Ollivier and Zurek \cite{Zurek} wrote a paper where they named this difference between the two ways of defining quantum mutual information quantum discord. They defined it slightly differently as they had the open system setting in mind, but I do not wish to enter any subtleties in this introductory article (an interested reader is encouraged to consult the review in \cite{Modi}). Physically, quantum discord, according to Zurek, represents the difference between the efficiency of classical and quantum Maxwell's demons, while in other interpretations it has also been linked to the fidelity of remote state preparation as well as to the difference in information extraction by local and global means (mathematically, at least, a protocol that is somewhat related to the Maxwell's demon interpretation). 

So discord seems to measure quantum correlations that go beyond just entanglement. Disentangled states can actually possess non-zero quantum discord. But is discord really a form of correlation? To answer that, we need to discuss an important property of any measure of correlation. 

One of the features of correlations is that they cannot increase by local operations (LO). If we do something to $A$ alone, and, independently, to $B$, we should not be able to correlate them to a higher degree than we started with. The intuition behind this is clear: we cannot correlate things more unless we are allowed to act on them jointly. Any separate action can only degrade the initial correlation (or, at best, preserve it). 

Both mutual information and entanglement are decreasing under LO (entanglement, in fact, under an even more general class, but this need not concern us here \cite{Vedral1}). This is actually straightforward to prove if we express both in terms of the quantum relative entropy $S(\sigma ||\rho)=tr (\sigma \log \sigma - \sigma \log \rho)$. Entanglement is then the relative entropy to the closest disentangled (separable) state \cite{Vedral1,Vedral2} (whose form is $\sum_n p_n \rho_A^n\otimes\rho_B^n$, where $p_n$ is any probability distribution), while the quantum mutual information is the relative entropy to the closest product states (which happens to be the product of the reduced states $\sigma_A\otimes\sigma_B$). Since quantum relative entropy is monotonic under completely positive maps and as the sets of product states and separable states are invarient under LO (which is a special set of completely positive maps), it follows that the quantum mutual information and entanglement are monotones (non-increasing) under LO (all the relevant proofs can be found in \cite{Vedral_RMP}).

However, monotonicity under LO is not true for discord! We can start with a state with no discord and actually create some by LO. A simple example is a stating state which is an equal mixture of $|00\rangle$ and $|11\rangle$, which can be converted by LO into a mixture of of $|00\rangle$ and $|1+\rangle$. It is therefore hard to think of discord as a form of correlations. Also, given that it can be created by local means, it is questionable if we can think of (all) discordant states as useful for quantum information processing. Having said this, there are examples of protocols where discord has an operational meaning \cite{Dakic,Gu}. It is also still an open question if universal quantum computation can be done without entanglmenet in the general case of mixed states. Maybe not all, but certainly some kind of discord could be of importance. 

This does not mean that we cannot quantify discord using the quantum relative entropy. We can take the relative entropy from a given state to the closest classically correlated state. This set, however, is not invarient under LOs which is why this measure fails to be a monotone (as examplified in the previous paragraph). 

\section{Outlook}

One should emphasise that though this article has dealt with bipartite systems only (for clarity, as well as for historical reasons), correlation measures can be generalized to many partite systems (see e.g. \cite{Amico} for entanglement in many-body systems and \cite{Modi} for discord and related measures). A way to do that is using the same relative entropy based logic outlines above (see also \cite{Modi2} for a unified view of all correlations based of the quantum relative entropy). 

Also, we did not discuss how we can tell if a given state has discord. The method is simple and it boils down to showing that correlations are non-vanishing in more than one basis \cite{Dakic2}. Classical correlations, according to this logic, are the ones that exist only in one basis (though this basis could be different for different subsystems, depending on how they couple to their environments, for instance). 

In conclusion, discord without entanglement can be seen as a form of classical correlation aided with quantum coherence (superpositions) at the level of individual subsystems. This is why the research on discord has naturally led to the research on quantifying quantum coherence.

\textit{Acknowledgments}: The author acknowledges funding from the John Templeton Foundation,
the National Research Foundation (Singapore), the Ministry of Education (Singapore), the Engineering
and Physical Sciences Research Council (UK), the Leverhulme Trust, the Oxford Martin
School, and Wolfson College, University of Oxford. This research is also supported by the National
Research Foundation, Prime Ministers Office, Singapore under its Competitive Research
Programme (CRP Award No. NRF- CRP14-2014-02) and administered by Centre for Quantum
Technologies, National University of Singapore.


\begin{thebibliography}{99}
%
\bibitem{Shannon} C. Shannon, The Bell System Technical Journal {\bf 27}, 379-423, 623-656 (1948).
%
\bibitem{Vedral_RMP} V. Vedral, Rev. Mod. Phys. {\bf 74}, 197 (2002).
%
\bibitem{Lindblad} G. Lindbad, Commun. math. Phys. {\bf 33}, 305 (1973).
%
\bibitem{Henderson} L. Henderson and V. Vedral, J. Phys. A: Math. Gen. {\bf 34}, 6899 (2001).
%
\bibitem{Zurek} H. Ollivier and W. H. Zurek, Phys. Rev. Lett. {\bf 88}, 017901 (2001).
%
\bibitem{Zurek2} W. H. Zurek, Phys. Rev. A {\bf 67}, 012320 (2003). 
%
\bibitem{Dakic} B. Dakic et al, Nature Physics {\bf 8}, 666 (2012).
%
\bibitem{Gu} M. Gu et al,  Nature Physics {\bf 8}, 671 (2012).
%
\bibitem{Modi} Kavan Modi, Aharon Brodutch, Hugo Cable, Tomasz Paterek, and Vlatko Vedral
Rev. Mod. Phys. {\bf 84}, 1655 (2012).
%
\bibitem{Vedral1} V. Vedral, M. B. Plenio, M. Rippin and P. L. Knight, Phys. Rev. Lett. {\bf 78}, 2275 (1997).
%
\bibitem{Vedral2} V. Vedral and M. B. Plenio, Phys. Rev. A {\bf 57}, 1619 (1998).
%
\bibitem{Modi2} K. Modi, T. Paterek, W. Son, V. Vedral, M. Williamson, Phys. Rev. Lett. {\bf 104}, 080501 (2010).
%
\bibitem{Amico} L. Amico, R. Fazio, A. Osterloh and V. Vedral, Rev. Mod. Phys. {\bf 80}, 1 (2008).
%
\bibitem{Dakic2} B. Dakic, V. Vedral and C. Brukner, Phys. Rev. Lett. {\bf 105}, 190502 (2010).


\end{thebibliography}
\end{document}